\documentstyle[preprint,prb,aps]{revtex}
\begin{document}

\draft
\preprint{IUCM94-015}

\title{Magnetic Oscillations and Quasiparticle Band Structure
in the Mixed State of Type-II Superconductors}

\author{M.R. Norman}
\address{ Materials Science Division, Argonne National Laboratory,
Argonne, IL 60439}
\author{A.H. MacDonald}
\address{Department of Physics, Indiana University,
Bloomington, IN 47405}
\author{Hiroshi Akera}
\address{Faculty of Engineering, Hokkaido University, Sapporo 060,
Japan}

\date{\today}

\maketitle

\begin{abstract}

We consider magnetic oscillations due to Landau quantization in
the mixed state of type-II superconductors.   Our work
is based on a previously developed formalism which allows
the mean-field gap equations of the Abrikosov state
to be conveniently solved in a Landau level representation.
We find that the quasiparticle band structure changes qualitatively
when the pairing self-energy becomes comparable to the
Landau level separation.  For small pairing self-energies,
Landau level mixing due to the superconducting order is weak
and magnetic oscillations survive in the superconducting state although
they are damped.  We find that the width of the quasiparticle
Landau levels in this regime varies approximately as $\Delta_0
n_{\mu}^{-1/4}$ where $\Delta_0$ is proportional to the magnitude
of the order parameter and $n_{\mu}$ is the Landau level index
at the Fermi energy.
For larger pairing self-energies, the lowest energy quasiparticle bands
occur in pairs which are nearly equally spaced from each other
and evolve with weakening magnetic field toward the bound states
of an isolated vortex core.  These bands have a weak
magnetic field dependence and magnetic oscillations vanish rapidly
in this regime.  We discuss recent observations of the
de Haas-van Alphen effect in the mixed state of several type II
superconductors in light of our results.

\end{abstract}


\section{Introduction}
\label{sec:intro}

For normal state electrons the study of
de Haas-van Alphen (dHvA) oscillations,
magnetic oscillations associated with Landau quantization in
a magnetic field, has proved
to be one of the most revealing probes of low-energy quasiparticle
excitations.\cite{shoenberg}  In the superconducting state, however, the
dHvA effect has not been studied systematically and
a great deal of confusion has surrounded its interpretation
when the effect has been observed.   An external magnetic field
is completely screened from the bulk of a superconductor
at sufficiently weak magnetic fields; this state of a superconductor
is known as the Meissner state.  For type-II superconductors a
second superconducting state in which the external field
is only partially screened, the Abrikosov\cite{abrikosov}
vortex lattice state or mixed state, occurs in stronger external
magnetic fields.  It is in this state that magnetic
oscillations can occur.  The first observation of
magnetic oscillations in the Abrikosov state
occurred nearly twenty years ago\cite{graebner} and interest
has been renewed recently with their observation in such materials
as $\rm{NbSe}_2$,\cite{onuki} $\rm{V}_3\rm{Si}$,\cite{mv3si,spring1}
$\rm{Nb}_3\rm{Sn}$,\cite{spring2} and in the high temperature superconductors
YBCO\cite{mybco,kido,haan} and BKBO.\cite{good}

Theoretically it was realized quite early\cite{hc2,hc2new,tesan1,nam,maniv}
that the upper critical magnetic field, $H_{c2}$,
above which the Abrikosov state
gives way to the normal state, shows magnetic oscillations
due to Landau quantization.  However much less progress has
been made on the question of what happens to dHvA
oscillations for fields substantially below
$H_{c2}$, principally because
of complications introduced by the broken translational
symmetry of the vortex lattice state.  Several authors
have addressed the limit where
pairing occurs within a single Landau level resulting in important
simplifications.\cite{akera,kumar,dukan,stephen,ryan}
This limit can pertain at low temperatures to fields just above the
semiclassical
$H_{c2}$ where the order parameter is very small.
(Plausible suggestions have been made\cite{rasolt}
concerning the possibility of reentrant superconductivity
for $H >> H_{c2}(T=0)$ but these still lack experimental
verification and considerations of the
vortex lattice state in the reentrant regime\cite{akera}
have been largely didactic.)
The works of Maki\cite{maki} and of Stephen,\cite{stephen2}
whose analyses are based on semiclassical
approximations for the electron Greens function\cite{semi}
in the mixed state, are more relevant to the intermediate field
situation.  Their results are based in part
on different versions of an approximation which involves
positional averaging over the vortex lattice
and was first introduced by Brandt {\em et al}.\cite{semi}
In this approach the vortex lattice acts much like a
random scattering potential which contributes to the
inverse lifetime of quasiparticle states without shifting
them away from the Fermi level.  When Landau quantization is
accounted for this scattering broadens the Landau levels
and therefore reduces the amplitude of the dHvA
oscillations.  An interesting alternate approach
was taken in a paper by Maniv
{\em et al},\cite{maniv} which studies magnetic oscillations in the
superconducting condensation energy.

In this paper we study magnetic
oscillations in the mixed state using a previously developed
formalism\cite{physicac} which allows the electronic structure of the Abrikosov
lattice state to be calculated conveniently in a Landau level basis.
This approach treats the effects of Landau quantization
without approximation within mean-field theory.
(A similar formalism was independently derived by
Rajagopal.\cite{rajaform})
Our conclusions are based in part on numerical calculations for
representative models of weak coupling superconductors and,
in part, on analytic perturbative calculations.
Our results differ qualitatively from those obtained in earlier work.
We find that magnetic oscillations in the normal state free-energy
and in the superconducting condensation energy have a strong
tendency to cancel, making studies of the condensation energy
alone misleading.  (The condensation energy is defined as the
difference between the free energies of normal and superconducting
states.) Close to $H_{c2}$, the off-diagonal self-energy
of the mean-field equations is small, and we find in agreement
with Maki\cite{maki} and Stephen\cite{stephen2} that the
quasiparticle bands consist of broadened Landau levels which are not
substantially shifted away from the Fermi level.  However we find that
the width of the Landau levels varies with the magnitude ($\Delta_0$)
of the order parameter and the Landau level index at the
Fermi level ($n_{\mu}$) as $ \Delta_0 n_{\mu}^{-1/4}$ rather than as
$ \Delta_0^2 n_{\mu}^{-1/2}$.  This qualitative difference is important
for the interpretation of experiments which show magnetic oscillations
in the mixed state and is a consequence of the degeneracy of the
Landau bands in the absence of superconducting order.
When the width of the Landau quasiparticle bands
becomes comparable to the Landau level separation we find that
there is a crossover in the quasiparticle spectrum.  The
lowest energy quasiparticle bands evolve into tight-binding
bands corresponding to the bound states of isolated vortices
and are insensitive to magnetic field strength.  The higher energy
quasiparticle bands retain Landau level character but are
shifted well away from the Fermi level.  Magnetic oscillations
are negligible in this regime.

Our paper is organized as follows.  In Section \ref{sec:formalism}
we briefly review our formalism for solving the non-linear gap equations
for the Abrikosov state in a Landau level representation.  Previous
calculations using this formalism focused on the self-consistently calculated
order parameter\cite{physicac} and the tunneling spectra.\cite{physicab}
Some practical details of these calculations are
discussed in Section \ref{sec:models}.
In Section \ref{sec:elstructure} we focus
on the quasiparticle electronic structure.
The non-linear gap equations are solved numerically for a
representative model of a weak-coupling superconductor.
dHvA oscillations are a consequence of the dependence
of the electronic structure on magnetic field, and specifically
result from Landau levels crossing through the chemical
potential as a function of magnetic field.
In Section \ref{sec:dhva} we discuss in detail the consequences for
magnetic oscillations of our results for the quasiparticle
spectrum in the mixed state.  Functional fits to the suppression of the
magnetic oscillations are derived and related to experimental data.
Section \ref{sec:variational}
contains a discussion of the magnetic field dependence of the
energy from a variational point of view which aims to explain
the importance of competition between pairing energy and kinetic
energy in reducing the amplitude of magnetic oscillations.
Finally in Section \ref{sec:conclusion} we briefly summarize our results.

\section{Non-Linear Gap Equations in the Abrikosov Lattice State}
\label{sec:formalism}

The discussion in this section and the illustrative calculations
in subsequent sections are for the case of two-dimensional
superconductors in perpendicular magnetic fields.
The restriction to two dimensions in this
section is simply a matter of notational simplicity; in the
models we use, the
quasiparticle states factorize into a planar part
which is explicitly exhibited below and a part associated with
the degree of freedom along the field direction.  Along the
field direction the pairing occurs between time-reversed
partners, $k_z$ and $-k_z$, as at zero magnetic field; including the
third direction necessitates merely the addition of a new index for the
quasiparticle states which must be summed over in constructing the
self-consistent off-diagonal self-energy.
The strength of the magnetic oscillations are stronger in the
two-dimensional case than they would be if we chose a three-dimensional
model; in subsequent sections, we have added comments at those points where
the translation to the three-dimensional case is not obvious.  The use of a
two-dimensional model to illustrate the points we wish to make
substantially simplifies our numerical calculations.  The three-dimensional
case is straightforward to treat, but is more computationally
demanding due to (i) having to treat a larger number of Landau parabolas
as opposed to discrete Landau levels and (ii) having to work with a larger
number of $\vec{k}$ points in a 3D wavevector space as opposed to a 2D one.

The effect of a magnetic field on electronic orbitals
in quantum mechanics appears through the vector potential
($\nabla \times \vec{A} = \vec{B}$).  For a non-zero average magnetic
field $\vec{A}$ is $\sim L$ where $L$ is the system size and,
unless the electronic states are localized, even a weak magnetic
field cannot be treated perturbatively.  A quantum treatment of
the electronic structure of the Abrikosov lattice state
must be performed in a basis
where the average magnetic field ($B_0$) is accounted for exactly; it
is this requirement which fundamentally changes the nature of
the mean-field equations.  To describe the vortex lattice
state it is convenient to use a Landau gauge ($ \vec{A}
= (0,B_0 x,0)$) in which the kinetic energy eigenstates are
\begin{equation}
\phi_{NX}(\vec{r}) = {1 \over \sqrt{L_y}}e^{-iXy/l^2}\phi_N ({x-X \over l})
\label{eq:2}
\end{equation}
where $ \phi_N (x) = (2^N N! \sqrt{\pi} l)^{-1/2}e^{-x^2 /2}H_N (x) $
and $H_N$ is a Hermite polynomial of order N ($ \ell ^2 = \hbar c/eB$).
The non-linear gap equations for non-uniform
superconductors,\cite{degennes} commonly known as the
Bogoliubov-de Gennes (BdG) equations,
are obtained by making a generalized Hartree-Fock
factorization of the equation of motion for electron creation and
annihilation operators.  To exploit
the translational symmetry of the vortex lattice state we first form
magnetic Bloch states from the $\phi_{NX}$:
\begin{equation}
\phi_{N\vec{k}} = \sqrt{a_x \over L_x} \sum_t e^{ik_x a_x t}e^{i\pi t^2 /4}
        \phi_{N,-k_y l^2 +ta_x}(\vec{r}).
\label{eq:1}
\end{equation}
where the sum is over all integers, and for general $a_x$ and $a_y$, one
restricts $k_y$ to an interval of length $ a_x/ \ell^2$ and
$k_x$ to an interval of length $ 2 \pi / a_x$.
(Note that there is one wavevector for each state in the Landau level.)
This basis is convenient when solving the gap equations for a
vortex lattice with primitive lattice vectors
$(0,a_y)$ and $(a_x,-a_y/2)$
where $a_x a_y = \pi /\ell^2$.  Solutions corresponding to
various vortex lattices can be found by choosing different values
for the vortex lattice aspect
ratio $R= 2a_x /a_y$ with $R= 1$ for a square lattice and $R= \sqrt{3}$
for the triangular lattice, which is the ground state
when Ginzburg-Landau theory applies.
We see below that in this representation only states with opposing
Bloch wavevectors
are coupled by the off-diagonal self-energy, as in the zero-field
mean-field equations.   Unlike the zero-field case,
however, the basis orbitals are labeled by wavevector and by
a Landau level index and the pairing self-energy is not
diagonal in the Landau level index.
The BdG secular matrix for the case of
singlet pairing is of dimension $2N_L$ where $N_L$ is the
number of Landau levels involved in the pairing:\cite{physicac}
\begin{equation}
(\xi_N - E_{\vec{k}}^{\mu}) u_{N\vec{k}}^{\mu} + \sum_M F_{\vec{k} NM}
         v_{M\vec{k}}^{\mu} = 0
\label{eq:4}
\end{equation}
\begin{equation}
(-\xi_N - E_{\vec{k}}^{\mu}) v_{N\vec{k}}^{\mu} + \sum_M F_{\vec{k} MN}^*
         u_{M\vec{k}}^{\mu}= 0.
\label{eq:5}
\end{equation}
Here $u_{N\vec{k}}^{\mu}$ is the coefficient of $\phi_{N\vec{k}}$ and
$v_{N\vec{k}}^{\mu}$ is
the coefficient of $\phi_{N-\vec{k}}^*$ in the real-space
Bogoliubov amplitudes $u(\vec{r})$ and $v(\vec{r})$ for the $\mu$'th solution,
and $\xi_N = \hbar \omega_c (N + 1/2) - \mu $ is the
kinetic energy of the $N-th$ Landau level measured from
the chemical potential.  ($\omega_c = e B / m c$ is the
cyclotron frequency.) $F_{\vec{k} NM}$ is the $\vec{k}$ dependent
pairing self-energy matrix in the Landau level representation and must
be determined self-consistently as we discuss below.

In deriving
these equations we have assumed that the magnetic flux density ($B$)
in the superconductor is uniform.  For
a strong type II superconductor, this approximation should be accurate
except for external fields close to the lower critical field.
The above equations also assume that we are dealing with an electron gas model
which is translationally invariant in the normal state; the
discussion could be modified to include band structure effects but
that complication is not discussed here.
As mentioned above,
Eqs.[\ref{eq:4},\ref{eq:5}] have been written in a form appropriate
to the two-dimensional case to avoid unnecessarily cluttering
the notation.  The illustrative calculations performed
in subsequent sections are for a two-dimensional electron gas
model with attractive interactions, since aspects associated with the
band structure of a particular material are not important for the
issues we plan to address and since magnetic oscillation phenomena
are more pronounced in the two-dimensional case.

We use a simple Bardeen-Cooper-Schrieffer (BCS)
model for the frequency dependence of
the attractive interaction giving rise to the pairing self-energy
so that the sums over Landau levels are restricted to states
whose kinetic energies lie within a cutoff
energy ($\omega_D$) of the chemical potential.
For a $\delta$-function attractive interaction the
off-diagonal self-energies are given by\cite{physicac}
\begin{equation}
F_{\vec{k} NM} = {-\lambda \hbar \omega_c \over 2 } \sum_j
\chi_{M+N-j}(\vec{k})
              D_j^{MN}\Delta_j
\label{eq:6}
\end{equation}
with
\begin{equation}
\chi_j (\vec{k}) = \sum_t e^{i2k_x a_x t}e^{-i \pi t^2 /2}\chi_j (2k_y l^2
             +2ta_x )
\label{eq:7}
\end{equation}
\begin{equation}
\chi_j(Y) = ({l \over \sqrt{2}})^{1/2}\phi_j ({Y \over \sqrt{2}l})
\label{eq:8}
\end{equation}
and
\begin{equation}
D_j^{NM} = ({j!(N+M-j)!N!M! \over 2^{N+M}})^{1/2} \sum_{m=0}^j {(-1)^{N-m}
       \over (j-m)!(N+m-j)!(M-m)!m!}
\label{eq:9}
\end{equation}
In these equations $\lambda$ is the BCS coupling constant
($\lambda \hbar \omega_c = V/(2 \pi \ell^2)$
where V is the strength of the attractive interaction).
The sum over $j$ in
Eq.[\ref{eq:6}] is over the possible partitionings of
the total quantized kinetic energy of the pair,
$\hbar \omega_c (N+M+1)$, into contributions from the
pair center of mass motion, $\hbar \omega_c (j+1/2)$, and
the pair relative motion, $\hbar \omega_c (N+M-j+1/2)$.

The derivation of this form for the pairing self-energy
closely follows the derivation\cite{hc2new} of the linearized gap equations
which is greatly simplified by making unitary transformations for
two-particle states between the representations in which
each particle has a definite Landau quantized kinetic energy
and the representation in which definite Landau quantized
kinetic energies reside in the center of mass and relative motion
of the pairs.  The $D_j^{NM}$, for which explicit expressions
are given in Eq.[\ref{eq:9}], are the matrix elements of the
unitary transformation.
$(D_j^{NM})^2$ is the probability that a pair of electrons
in Landau levels $N$ and $M$ will have center-of-mass kinetic
energy $\hbar \omega_c (j+1/2)$.  $\Delta_j$ in Eq.[\ref{eq:6}]
is the amplitude for electron pairing with center-of-mass kinetic energy
$ \hbar \omega_c (j+1/2)$ and is directly related to the Landau level
expansion of the Ginzburg-Landau order
parameter.\cite{degennes,tinkham}  In our formalism, the self-consistent
solution of the BdG equations reduces to the self-consistent
determination of these parameters.  In practice it is only
necessary to determine a small number of parameters.  This is
particularly true near $H_{c2}$ where Ginzburg-Landau theory
tells us that only $\Delta_0$ (the Abrikosov solution) is significantly
different from
zero and we will discuss the approximation where only this single
parameter is retained at length in Section \ref{sec:variational}.
Even at weaker magnetic fields, it follows from symmetry considerations
that for a triangular Abrikosov lattice, $\Delta_j$ is different from
zero only if $j$ is a multiple of six and this property helps to
keep the number of parameters which need to be determined
self-consistently small.  The higher $j$ components of
$\Delta_j$ are essential, however, in describing the reduction
of the vortex core size compared to the lattice constant
of the vortex lattice as the field is reduced.  $\Delta_j$ is
given in terms of the eigenvalues and eigenvectors of the
Landau representation BdG equations by the following
equation:\cite{physicac}
\begin{equation}
\Delta_j = -\sum_{NM} D_j^{MN} \sum_{\vec{k}} {4 \pi l a_x \over L_xL_y}
    \chi_{M+N-j}^*(\vec{k}) \sum_{\mu} (1-f_{\vec{k}
    \uparrow}^{\mu}-f_{\vec{k} \downarrow}^{\mu}) u_{N\vec{k}}^{\mu}
    v_{M\vec{k}}^{*\mu}
\label{eq:12}
\end{equation}
where $E_{\vec{k}}^{\mu}$ is the
$\mu$'th positive eigenvalue of the Landau representation BdG equations,
and $f_{\vec{k} \sigma}^{\mu}$ is the Fermi function evaluated at the
appropriate quasiparticle energy.   If Zeeman coupling is included
the Fermi functions should be evaluated at energies,
\begin{equation}
E_{\vec{k} \sigma}^{\mu} = E_{\vec{k}}^{\mu} - g^*\hbar \omega_c \sigma /4
\label{eq:13}
\end{equation}
where $\sigma = \pm 1 $ is the spin index and $g^*$ the effective
g-factor $(gm^*/m)$.

The diagonal order parameter in a position representation is related
to the $\Delta_j$ by
\begin{equation}
\Delta(\vec{r}) = {(\sqrt{2}L_y l)^{1/2} \over 4 \pi l^2} \sum_j (-1)^j
      \Delta_j \sum_t e^{i \pi t^2 /2} \phi_{j,\sqrt{2}ta_x}(\sqrt{2} \vec{r})
\label{eq:14}
\end{equation}
Once the BdG equations have been evaluated self-consistently at
a particular magnetic field, the free energy can be calculated
using\cite{physicac}
\begin{equation}
F = \sum_{N\sigma} \xi_{N\sigma}N_{N\sigma} + E_P - TS
\label{eq:15}
\end{equation}
where the pairing self-energy is
\begin{equation}
E_P = -\lambda \hbar \omega_c {L_x L_y \over 8a_x \pi l} \sum_j |\Delta_j|^2
\label{eq:16}
\end{equation}
Here $N_{N\sigma}$ is the occupation number of the $N\sigma$ Landau level
\begin{equation}
N_{N\sigma} = \sum_{\mu\vec{k}} f_{\vec{k}\sigma}^{\mu}|u_{N\vec{k}}^{\mu}|^2
 + (1-f_{\vec{k}\bar{\sigma}}^{\mu})|v_{N\vec{k}}^{\mu}|^2
\label{eq:17}
\end{equation}
and $S$ is the entropy
\begin{equation}
S = -k_B \sum_{\mu\vec{k}\sigma} (1-f_{\vec{k}\sigma}^{\mu})
\ln(1-f_{\vec{k}\sigma}^{\mu}) + f_{\vec{k}\sigma}^{\mu}\ln
f_{\vec{k}\sigma}^{\mu}
\label{eq:18}
\end{equation}

\section{Model Calculations}
\label{sec:models}

In this section we discuss some practical details of the
model calculations performed using the above formalism
and present some results for the magnetic field dependence
of $T_{c2}$ and the zero-temperature order parameter.
This results serve as a characterization of the model for
which we will study magnetic oscillations.

We have found that it is necessary to take some care in cutting
off the attractive interaction of the model away from the Fermi energy.
For many purposes it is adequate to
treat the cut-off in the BCS fashion, that is to simply set the
attractive interaction to zero when the difference between the
normal state quasiparticle energy and the chemical potential exceeds
some value.  This procedure can lead to undesirable consequences.
In our own work on the
linearized gap equation,\cite{hc2new} we found that that it leads to
unphysical features in $H_{c2}$ associated with Landau levels
passing through the cut-off energy.  In the work of
Markiewicz {\em et al},\cite{mark} strong features were found
in the gap function of the same origin.  In our preliminary
work on this problem, we found a more severe problem.  In the two-dimensional
sharp cut-off model with fixed chemical potential,
we found that the superconducting
condensation energy increases smoothly
with increasing field as the Landau level degeneracy increases
and then decreases discontinuously as the the number
of Landau levels within the pairing cut-off decreases by one.
This causes the typical magnetization to be paramagnetic (since the
critical temperature increases with field\cite{csdm}) with
sharp diamagnetic spikes at particular fields.
By going to a smooth cut-off, though, we recover the continuous
diamagnetic behavior we expect to be caused by the reduction of
superconducting condensation energy in a magnetic field.
A pragmatic solution is achieved by using a model where the pairing
interaction between Landau levels $N$ and $M$ is scaled by
$\sqrt{W_N W_M}$ where
\begin{equation}
W_N = 1.55 e^{-(\xi_N/0.5\omega_D)^4}.
\label{eq:19}
\end{equation}
The sums are still restricted to energies within $\omega_D$ of the
chemical potential but the interactions at the `edge' of the pairing
window are sufficiently weak that they do not give rise to
significant anomalies.  We have used this procedure in all our
calculations.

We note that because we are using full quantum wavefunctions for the
Landau levels, we are currently limited to single particle Landau indices
of 60 or less. This practical limit is reached because of numerical
difficulties originating in
the rapid oscillations of the wavefunctions for large Landau index.
This problem could be circumvented
by using semiclassical approximations when the Landau level indices are
too large.

We have chosen to work in the grand-canonical
ensemble where the chemical potential rather than the particle number
is held fixed.  The difference between the dependence of the
magnetization on density and the dependence of the magnetization on
chemical potential has been extensively discussed by
Shoenberg\cite{shoen2} for the case of the normal state.

For every model we have checked we find that when
the external magnetic field is weaker than
the semiclassical upper critical field, the
energy is minimized by a triangular vortex lattice in agreement
with Ginzburg-Landau theory.  (This is not the case for solutions
of the mean-field equations at stronger external fields
which are in the reentrant quantum regime.\cite{physicac,physicab})
All the model calculations in subsequent sections of this
paper are for triangular flux lattices and one can
exploit the symmetries of that situation.  It is possible to show that
for a triangular vortex lattice the BdG equations have the
symmetry of a triangular Brillouin zone with an origin at
$\Gamma \equiv (\frac{\pi}{4a_x},\frac{\pi}{2a_y})$.
This allows us to limit our sums over wavevectors to
$\vec{k}$'s in an irreducible triangle of the zone
with vertices at
$\Gamma$, $ M \equiv (\frac{3\pi}{4a_x},\frac{\pi}{2a_y})$ and
$K \equiv (\frac{3\pi}{4a_x},\frac{\pi}{6a_y})$.  The area of this
irreducible triangle is $1/12$ of the area of the full zone
and the labeling of the points
at the vertices is the conventional one for symmetry points in a
triangular lattice.  For most of our calculations, we have used a
$\vec{k}$ mesh with a spacing which is $0.2$ times the
$\Gamma-M$ distance for the wavevector sums, although we have done
a number of calculations with a mesh which is twice as fine.
The two meshes lead respectively to $21$ and $66$ $\vec{k}$'s
in the irreducible triangle.

Most of the self-consistent calculations in subsequent sections
used a coupling constant, $\lambda =0.75$.
This choice places the semiclassical upper
critical field at a low enough Landau index (strong enough
magnetic field) so that we could access most
of the semiclassical phase region with our code's restriction on the maximum
Landau level index.  A large ratio of the cut-off, $\omega_D$, to the chemical
potential, $\mu$, of $\frac{1}{2}$ was chosen so as to maximize the number of
Landau levels involved in the pairing.  For the sake of definiteness
we choose the g-factor to be zero in our model calculations.
(Zeeman spin-splitting does modulate dHvA oscillations in the
normal state in a way which is well understood\cite{shoenberg}
and will have a similar effect in the superconducting state.)

In Fig. 1, we show a plot of the critical temperature,
$T_{c2}$, versus field, for this model.   The field is
parameterized in terms of $n_{\mu} \equiv \mu/ \hbar \omega_c -1/2$ which
gives the Landau level index of the Fermi level.
(Note that, in contrast to our previous work,\cite{hc2new} no magnetic
oscillations are visible
in this curve at larger values of $n_{\mu}$;
the oscillations at large $n_{\mu}$ found previously are an artifact of using
a sharp cut-off.)  At strong enough fields, the critical temperature
is driven to a low enough value that $k_B T_{c2}$ is much smaller
than $\hbar \omega_c$ and the system crosses over
to the quantum regime where magnetic oscillations in $T_{c2}$ appear.
In this regime a significant portion of the pairing occurs
within the Landau level at the Fermi level, even at $T_{c2}$.
These quantum oscillations in $T_{c2}$ are much less
robust in three-dimensional models and do not appear
until $ k_B T_{c2}  < \hbar \omega_c / n_{\mu} $.
Zeeman spin-splitting and broadening of Landau levels due to
disorder also will suppress these oscillations.\cite{nam}
The maxima in $T_{c2}$ occur when the density of states at the
chemical potential is largest; {\em i.e.,} when a
Landau level occurs at the chemical potential.  Regions where
$T_{c2}$ vanishes occur because we work at fixed chemical potential;
at fixed particle number, the chemical
potential is locked to a Landau level, so cusps occur instead of gaps.
For the present model once $n_{\mu} > 12$ no quantum effects
can be seen in the dependence of $T_{c2}$ on field.

Also shown in Fig. 1 is a plot of the root-mean-square spatial average of the
off-diagonal self-energy in a coordinate representation
($F \simeq 0.44 \lambda \hbar \omega_c \Delta_0$)
for this model obtained by solving
the non-linear gap equation self-consistently for the same model
at $k_BT/\mu=10^{-3}$.  As we approach the weak-field limit we expect that
$\Delta (\vec{r})$ should be constant over much of the area of the
system and that $F$ should approach the zero-field BCS energy gap
for this model.  We see in Fig. 1 that at weaker fields the ratio of
$F$ to $T_{c2}$ approaches a constant.
The ratio of these quantities varies from $1.87$ for $n_{\mu}=20$
to $1.77$ for $n_{\mu}=40$, compared to the zero-field
BCS value of $1.76$.  Magnetic oscillations occur
in this curve since the temperature used is low.
Their suppression at weaker magnetic fields must originate
from changes in the electronic structure associated with
superconductivity.  It is this suppression which is our primary
interest in the present paper.

In Fig. 2, we show the spatial dependence of the self-consistent
order parameter at $k_BT/\mu=10^{-3}$ along the line connecting neighboring
vortex cores.
Results are shown for various fields from
$n_{\mu} = 12$  to $n_{\mu} =39$.  At small values of $n_{\mu}$
({\em i.e.} near $H_{c2}$) the order parameter is close to the Abrikosov
order parameter which is obtained when only $\Delta_0 \ne 0$.
At larger values of $n_{\mu}$, the vortex core size becomes small
compared to the distance between vortex cores
and the magnitude of the order parameter becomes nearly constant
outside the cores as expected. (Many $\Delta_j$ need to be non-zero
to obtain this behavior.)  Interestingly, the dependence on $r$ has
oscillations which, as we will see later, are related to discrete
quasi-bound states in the vortex cores.

\section{Flux Lattice Quasiparticle Bands}
\label{sec:elstructure}

In this section we look in detail at the quasiparticle
band structure in the mixed state.  In Figs. 3 and 4, we show plots of
$E(\vec{k})$ along high symmetry directions in the
irreducible triangle of the Brillouin zone
for the ten lowest energy quasiparticle bands at
a field close to the semiclassical
$H_{c2}(T=0)$ ($n_{\mu}=15$) and a much weaker magnetic field
($n_{\mu}=40$).  From Fig. 3 it is
clear that the quasiparticle bands at $n_{\mu} = 15$ can
be regarded as broadened Landau level bands.   The quasiparticle
bands in the normal state are dispersionless and when $n_{\mu}$
is an integer one band exists at zero energy and two bands at
each multiple of $\hbar \omega_c$ (particle, hole).
dHvA oscillations
are appreciable when the band at the Fermi energy is broadened
by less than $\hbar \omega_c$ and we see in Fig. 3 that this condition
is still satisfied at $n_{\mu} = 15$.  Moreover, zero energy
quasiparticles, which contribute strongly to
the magnetic oscillations,
occur at several $\vec{k}$ points in the zone as can be seen in Fig. 3.

As $n_{\mu}$ increases and the order parameter gets larger the
quasiparticle bands continue to broaden, but then cross over to a regime
where the lowest energy bands shift away from zero and begin to narrow.
In this regime, the eigenvalues have a weak dependence on field.  This
can be seen in Fig. 5 where the average energy of each pair of bands is
plotted in chemical potential units versus $n_{\mu}$.  At low $n_{\mu}$,
the decrease of the eigenvalues in field is just a reflection of the
decrease in cyclotron frequency with decreasing field ({\em i.e.,} the levels
behave like Landau levels).  These levels then cross over to field
independent behavior at high $n_{\mu}$ with an energy scale characteristic
of vortex core bound states.  In fact,
the low energy quasiparticle bands shown in Fig. 4 occur in pairs
and the energies of the lower energy member of the three lowest pair of bands
at the $\Gamma$ point
are approximately in the ratio of 1:3:5.  This is the ratio of
energies expected for the lowest energy quasiparticle states bound to
an isolated vortex core and in fact the lowest energy eigenvalue is close
to the predicted\cite{core} value of $\Delta^2/\mu$.  We believe that the
low energy
bands seen in Fig. 4 correspond to quasi-bound quasiparticles which
can tunnel from core to core, with a weak field dependence of the energies
since the cores have a size much smaller than a magnetic length (see Fig. 2).
There are two bands for each
bound state since each band has one state for each electron
magnetic flux quantum through the system
($\Phi_e = h c /e$) while the system has one vortex
for each superconducting magnetic flux quantum ($\Phi_{sc} = \Phi_e /2$),
{\em i.e,} there are two vortices for each single particle orbital
in the normal state Landau levels.  There is also a connection of this
doubling behavior to the normal state where pairs of
excitations (particle, hole) exist, in that if one member of a
pair has weight in Landau level n for the $u$ component of the wavefunction,
the other has approximately the same weight in level n-1 for the $v$ component.
The difference is that at low $n_{\mu}$, the weight is primarily in one Landau
level whereas at high $n_{\mu}$, the weight is distributed among many Landau
levels.

This vortex core interpretation of the low energy bands at $n_{\mu} = 40$
has been verified by further analyzing the corresponding wavefunctions.
Near an isolated vortex core,\cite{core} the quasiparticle
amplitudes can be expanded in a set of basis wavefunctions
with definite angular momentum.  The lowest energy bound
quasiparticle state is formed
from pairs with angular momentum $0$ for the $u$ component
and angular momentum $-1$ for the $v$ component; the next bound
quasiparticle state is formed from pairs with angular momentum
$-1$ for the $u$ component and $-2$ for the $v$ component and so on.
The order parameter is proportional to the product $uv^*$.  The
contribution of the $m$-th bound state to the order parameter
is therefore proportional to $r^{2m+1} \exp ( i \phi)$ at small
$r$ where $r$ and $\phi$ are circular coordinates in the system
centered on the vortex core.  The common angular dependence is
required to self-consistently maintain
the unit vorticity of the vortex.  In Fig. 6, we plot the contributions
to the order parameter from the lowest two quasiparticle bands
for $n_{\mu} = 40$.  We see that the contribution from the lowest
pair of bands is proportional to $r$ at small $r$ while that
from the second pair is proportional to $r^3$, as expected from
the above discussion.   Near the vortex core the order parameter is
dominated by the contribution from the lowest energy pair of
quasiparticle bands; far from the vortex core the low energy
`bound-state' bands contribute little to the order parameter.
The non-monotonic behavior of the order parameter as a function
of $r$ seen in Fig. 2 is due to the `shell structure' of the
quasi-bound states.

When $n_{\mu}$ is close to an integer and $\Delta$ is small
the BdG equations can be truncated to the Landau level closest to
the Fermi level.  The lowest energy quasiparticle band dispersion
in this quantum limit is given by:\cite{akera,kumar,dukan,stephen}
\begin{equation}
E(\vec{k}) = \sqrt{ \xi_N^2 + |F_{\vec{k} N N}|^2}.
\label{eq:qlimit}
\end{equation}
Since $F_{\vec{k} N N}$ generically vanishes at some value of
$\vec{k}$ this implies gapless behavior in the quantum limit
in our two-dimensional model when $\xi_N=0$.
For the hexagonal lattice this approximation leads to a
third order zero in the excitation spectrum at the $\Gamma$ point and a first
order zero at the K point, as pointed
out by Dukan {\em et al}.\cite{dukan}  We also find five other points
along the three symmetry lines of the zone where the gap goes to zero.
For three-dimensional models\cite{dukan} $\xi_N$ depends on $k_z$ and
there will always be a $k_z$ for which $\xi_N$ vanishes for each
occupied Landau level in this approximation.  As pointed out recently by
Dukan {\em et al}\cite{dukan} gapless behavior is not ruined by
the weak Landau level mixing which is always present and persists
for a finite range of magnetic field below $H_{c2}$.  In
Fig. 7 we plot the lowest two quasiparticle energies at
21 inequivalent $\vec{k}$ points as a function of $n_{\mu}$.
Near the quantum limit, gapless excitations occur for integer
values of $n_{\mu}$.  At weaker fields gapless excitations occur
in our two-dimensional model at more widely spaced and,
in general, non-integral values of $n_{\mu}$.  (Presumably for
a three-dimensional model, there would still be a discrete set of $k_z$ values
at which gapless excitations occur in this regime.)
The eigenvalue spectrum in the quantum limit ({\em i.e.,} diagonal)
approximation
differs significantly from what the full theory gives and overemphasizes
gapless
behavior.  (It is the terms off-diagonal in Landau level index which
are responsible for the decrease of $T_{c2}$ with increasing field, so one
can never ignore them in the semiclassical regime.)  Nevertheless, true gapped
behavior sets
in only when the superconducting order is sufficiently
strong to increase the energies of the
quasi-bound states in the vortices above $\hbar \omega_c$.
We have been unable to divine any simple principles behind the
seemingly complex pattern of eigenvalues in the crossover
region between the quantum limit regime and the vortex core
bound state regime.  It appears that some kind of non-monotonic
and possibly oscillatory behavior occurs which does not originate
from Landau quantization.  This behavior is characterized by oscillations
between wide and narrow bands, with the lowest energy state of the wide bands
tending to alternate between the $\Gamma$ and $K$ points of the zone.  This
behavior is reminiscent of a tight-binding effect, indicating that the source
of this `long period' behavior is vortex-vortex interactions ({\em i.e.,} due
to the spatial modulation of the order parameter).  At the $\Gamma$ point there
is a pattern of oscillations which are approximately periodic in $n_{\mu}$ with
a period roughly of
six which reflect the fact that at this high symmetry point in
the irreducible triangle only Landau levels with
indices differing by six are mixed in the $u$ amplitude of
the quasiparticle state.  The $v$ amplitude behaves in the same
way with Landau level indices which for one member of the pair of bands is
offset by $1$ and the other member by $3$
from the $u$ amplitude Landau level indices.  The field dependencies
at lower symmetry $\vec{k}$ points are less simple in this
regime and there is a complicated pattern of avoided level crossings.
We have examined this regime for several different models and
have not succeeded in identifying universal features in the electronic
structure in the crossover regime except as mentioned above.
In Fig. 8 we show results for the field dependence of
the quasiparticle spectrum for the model with $\lambda$=0.55.
For this case, one crosses over from the quantum regime to the
semiclassical regime for $n_{\mu} \sim 21$.
No apparent long period structure is seen in the crossover regime.
For this model quasi-bound states just begin to emerge as we reach
our code's limitations at $n_{\mu} \sim 40$.

In Fig. 9 we show a plot of the Landau level occupation numbers
in the semiclassical regime at $n_{\mu} = 40$.
In the zero-field BCS case, it is known that the momentum
distribution function at T=0 is approximately that of a Fermi function at
$T_c$.\cite{tinkham}  The analogous statement applies in the
semiclassical mixed-state regime, as demonstrated in Fig. 9.
A least squares fit of the Landau level occupation numbers to
a Fermi function parameterized by a temperature $T$
gives an optimal $T$ equal to $T_{c2}$ to within a few percent.

\section{de Haas-van Alphen Oscillations}
\label{sec:dhva}

To determine the magnetization M in the grand-canonical ensemble
we self-consistently solve the mean-field equations at a set of
fields spaced so that $\Delta n_{\mu} =0.1$ and evaluate
the free energy using Eq.[\ref{eq:15}].  A finite difference
approximation is used for the derivative with respect to field
which yields a two-dimensional magnetization we define by
\begin{equation}
M_{2D}(B) =  - \frac{\partial F}{\partial B}.
\label{eq:magn}
\end{equation}
For the normal state the resulting magnetization accurately reproduces
the analytically known $T=0$ result\cite{shoenberg} in which $M_{2D}/(N \mu_B)$
is periodic in $n_{\mu}$ with period one and varies between $-1$ and $1$.
The magnetization in the normal state
jumps by $2 N \mu_B$ whenever the chemical potential
passes through one of the degenerate Landau levels.  It is the bunching
of the density of states into Landau levels which leads to the period
one oscillations, known as dHvA oscillations.
We remark that in evaluating the magnetization we have ignored the
screening of the external magnetic field so that our approach is valid
only for strong type-II superconductors.  For a three-dimensional
system composed of isolated layers, the bulk magnetization
is related to the two-dimensional magnetization defined above by
$M = M_{2D}/Ad$ where $A$ is the area of the two-dimensional system
and $d$ is the separation between two-dimensional layers.  Our
approximation is valid (using Gaussian units) as long as $M << B$.

In Fig. 10 we show a plot of $M_{2D}(H)$ for the $\lambda=0.75$ model
at a very low temperature.
At small $n_{\mu}$ (strong field), there are dHvA
oscillations with period one as in the normal state.  The dHvA
oscillations are rapidly damped
as the superconducting order develops.  Comparing with Fig. 7 we see
that strong dHvA oscillations are present only in
the quantum regime where the quasiparticle bands retain a clear
Landau level structure.  There is some structure in the magnetization
curve in the complicated crossover regime discussed above but these
features are {\em not} dHvA oscillations and we suspect
that they will tend to be washed out in three-dimensional models.
At weaker fields ($n_{\mu} > \sim 22$) we cross over to the regime
where bound states are supported by the vortex core and the field
dependence of the magnetization becomes relatively featureless.
The negative sign of the magnetization in this region
occurs because the
condensation energy associated with superconducting order
decreases with magnetic field.  If the field dependence of these
magnetization curves is Fourier transformed with respect to
$n_{\mu} \propto 1/H$, the peak at period one comes almost
entirely from the quantum regime.  At higher temperatures where the
magnetic oscillations are damped out, we find that the
magnetization is proportional to $1/H - 1/H_{c2}$ in contrast to
Ginzburg-Landau theory where it is proportional to $H - H_{c2}$.

In Fig. 11, we plot the contributions to $M_{2D}$ from the
kinetic energy and the pairing contributions to the magnetization
at zero temperature.  We see that in all regimes the two
contributions to the magnetization oppose each other.
This should be expected since the superconducting state
is always formed at a cost in kinetic energy in order
to take advantage of the attractive interparticle interactions.
This occurs in both quantum and semiclassical limits.
The quantum limit is particularly simple.  Pairing is strongest
when $n_{\mu}$ is an integer leading to a maximum in both the
increase of kinetic energy due to superconducting order
and in the pairing self-energy.  The total free-energy of the
system is decreased by much less than either component is shifted.
The total free-energy in the normal state is at a maximum when
$n_{\mu}$ is an integer and at a minimum when $n_{\mu}$ is at
a midpoint between integers; at such values of $n_{\mu}$ the
free energy is identical to the free energy in the absence of
a magnetic field.  The difference between minimum and
maximum free energies is reduced by introducing superconducting order
so that the amplitude of the dHvA oscillations is
reduced because of superconductivity.  At the same hand the amplitude
of the oscillations in the kinetic energy alone is increased.
This behavior reveals the importance of determining the superconducting
order self-consistently within each dHvA oscillation
period.  Earlier work which neglected dHvA oscillations
in the superconducting order itself will not be valid in general.
Because of self-consistency, oscillations in the superconducting
condensation energy also tend to oppose oscillations in the normal state
energy.  Well outside of the quantum regime this cancellation is nearly
complete.  Therefore the interesting work of Maniv
{\em et al}\cite{maniv} which focused on oscillations in the
condensation energy alone also has limited validity.
We comment further on self-consistency and the cancellation between
normal state and condensation energy magnetic oscillations in the
following section.

In Fig. 12, we show $M_{2D}(H)$ for the model with $\lambda = 0.55$.  Again,
magnetic oscillations are damped in the quantum regime and
exhibit a complicated pattern in the crossover regime which is
associated with the appearance and disappearance of gapless
behavior as a function of field.  As stated previously we believe
that these oscillations will wash out for three-dimensional models.
There is no simple relationship
between the magnetizations in the crossover regime for
the $\lambda = 0.55 $ and $\lambda = 0.75 $ models.   For the
$\lambda = 0.55$ model, the regime of quasi-bound vortex
states is just approached at the weakest
magnetic fields we have considered.

Our results for these models cannot be compared directly
with experiments on specific materials.  However it is
possible to make some general qualitative remarks that we believe are
important for the interpretation of experiments.  In the
Lifshitz-Kosevich theory\cite{lktheory} of dHvA
oscillations in the normal state the amplitude of the oscillations
is reduced by a factor of
\begin{equation}
R_0 = \exp (- \pi / \omega_c \tau_0)
\label{eq:r0}
\end{equation}
because of disorder broadening of Landau levels.  This specific
result is based on the assumption of a magnetic field independent
Lorentzian lineshape, with a full width at half maximum of $\hbar /\tau_0$,
for the density of states of the disorder broadened Landau level.
In the superconducting state (with no disorder)
the density of states of a broadened Landau level is not
Lorentian and, more importantly, it is not independent of
magnetic field.  Within each period of the dHvA
oscillation the superconducting order is strongest and the Landau level
is broadest when $n_{\mu}$ is an integer.  This Landau level broadening
reduces the increase in free-energy due to Landau quantization and
therefore reduces the amplitude of the dHvA oscillations.
Following Springford's group\cite{spring1,spring2} we therefore assume that the
damping of dHvA oscillations due to superconductivity
in the mixed state is given approximately by
\begin{equation}
R_S = \exp ( -\pi /\omega_c \tau_S)
\label{eq:rs}
\end{equation}
where $\hbar/ \tau_S$ is some measure of the
Landau level width in the mixed state of the $n_{\mu}$'th Landau level
when $\xi_{n_{\mu}} = 0$.  (In the case of three-dimensional models,
dHvA oscillations come from $k_z$ values where the
Landau level index at the Fermi level is a local minimum or maximum
(extremal orbits); the remarks below apply as well to the three-dimensional
case.)

In the region near $H_{c2}$ where
dHvA oscillations are strong we can assume that only
$\Delta_0 \ne 0$ so that from Eq.[\ref{eq:qlimit}]
\begin{equation}
\hbar \tau_S^{-1} \propto \lambda \hbar \omega_c \Delta_0 n_{\mu}^{-1/4}.
\label{eq:taus}
\end{equation}
The factor $n_{\mu}^{-1/4}$ which appears in Eq.[\ref{eq:taus}]
comes from the large quantum number limit\cite{ajp} of the factor
$D_0^{n_{\mu}n_{\mu}}$ in the expression for the off-diagonal self-energy;
$(D_0^{n_{\mu}n_{\mu}})^2$ is the probability\cite{ajp} for two electrons with
Landau level index $n_{\mu}$ to have center-of-mass kinetic energy $\hbar
\omega_c/2$ (Abrikosov solution) and its rate of decrease with $n_{\mu}$ in
the large quantum
number limit can be understood from simple phase-space considerations.
The actual dispersion relation within the quasiparticle Landau band
is complicated for large Landau indices
and we estimate the proportionality constant in Eq.[\ref{eq:taus}]
by numerically calculating the standard deviation of quasiparticle
energies within the band.  Results are shown in Fig. 13 from which
we infer that
\begin{equation}
\hbar \tau_S^{-1} \sim 0.15 \lambda \hbar \omega_c
\Delta_0 n_{\mu}^{-1/4}.
\label{eq:width0}
\end{equation}
As emphasized above $\Delta_0$ is itself
a strongly oscillating function in the quantum limit so that some
caution must be used in applying this relationship.  An exception
occurs for dHvA oscillations coming from a small piece
of Fermi surface in a system where the field dependence of $\Delta_0$
is dominated by larger pieces of Fermi surface.  In this case it may
be possible to use the dHvA oscillations from a small
piece of Fermi surface as a probe to look at the field dependence of
$\Delta_0$ as suggested recently by Harrison {\em et al}.\cite{spring2}

We can use these results to estimate the typical range of fields over which
dHvA oscillations from large pieces of Fermi surface
will be observable.  As mentioned in Sec. \ref{sec:models} the spatial
average of the off-diagonal self-energy $F \sim 0.44 \lambda \hbar
\omega_c \Delta_0$ retains its familiar BCS relationship
to the critical temperature for almost all fields.  Assuming this
relationship, we obtain
\begin{equation}
\hbar \tau_S^{-1}
\sim 0.6 n_{\mu}^{-1/4} k_B T_{c2}
\label{eq:width1}
\end{equation}
Using a Ginzburg-Landau form for the field dependence of
$T_{c2}$\cite{degennes}
this becomes
\begin{equation}
\hbar \tau_S^{-1}
\sim 0.6 n_{\mu}^{-1/4} k_B T_c \sqrt{1-H/H_{c2}}.
\label{eq:width2}
\end{equation}
dHvA oscillations will be strongly damped when the right
hand side of Eq.[\ref{eq:width2}] is larger than $\hbar \omega_c$.
Since the typical value of $\hbar \omega_c$ at $H_{c2}$ is\cite{ajp}
$\sim (k_B T_c)^2/ \mu$ this occurs for $(H_{c2}-H)/H_{c2} > k_B T_c /\mu$
which is a number much less than one.  (We have used $n_{\mu} \sim
\mu / (\hbar \omega_c)$ to obtain this estimate.)
For our simple model $k_B T_c/ \mu \sim 0.1$
so that this simple estimate is consistent with the range
of fields over which we see dHvA oscillations persist.
For more realistic models, $k_B T_c / \mu $ is typically much smaller,
at least for the largest pieces of Fermi surface, and dHvA
oscillations should be seen only close to $H_{c2}$.  We remark that
Eq.[\ref{eq:width2}] implies that small pieces of the Fermi surface
are more damped than
large ones due to the $n_{\mu}^{-1/4}$ factor.  Finally, we note that
magnetic oscillations of the gap function are determined by the large
pieces of the Fermi surface, and so the gap function will oscillate many
times over the period of a small orbit.
This could potentially complicate the analysis of such orbits.

Our results differ qualitatively from those of Maki\cite{maki} and
Stephen\cite{stephen2} who find, in our notation,
that $\hbar \tau_S^{-1} \sim |F_{NN}|^2/ (\hbar \omega_c)$ rather
than $\sim |F_{NN}|$.  In calculating the Landau level width these
authors do not take account of the Landau level quantization of the
density of final states into which the quasiparticles can be scattered.
The Landau level width can then be calculated treating the scattering
of quasiparticles by the vortex lattice using what is essentially a
golden rule.  In this language we find that
the density of states in the quasiparticle Landau level is
enhanced over the zero-field value
by a factor of $ \omega_c \tau_S$.  The density of final states in the
golden rule calculation is inversely proportional to $\hbar \tau_S^{-1}$
so that we find $\hbar \tau_S^{-1} \sim |F_{NN}|^2 / \hbar \tau_S^{-1}$.
When the Landau level is already broad because of disorder
scattering ($ \omega_c \tau_0 << 1 $), which we have neglected here,
the results of Maki\cite{maki} and Stephen\cite{stephen2} are more
appropriate than ours.  However, in this case dHvA
oscillations will be difficult to observe even in the normal state
and we believe that our result will typically be more useful in
analyzing experimental results.  This discussion underscores the
serious approximations involved in assuming that disorder and
superconducting order are responsible for additive contributions
to the Landau level width and correspondingly to independent
attenuation factors in the expression for dHvA oscillation
amplitudes.  We have focused here on the situation where the
dominant attenuation factor is that due to the presence of
superconducting order since it is in this situation that
dHvA oscillations have the greatest potential as a probe
of the mixed state.

As a further test of these ideas, we have analyzed the Fourier transform
of our calculations in more detail.  The problem with analyzing the
magnetization presented in Figs. 10 and 12 is that single period oscillations
exist only over a small field range due to the rapid growth of the order
parameter below $H_{c2}$ and the limited number of Landau levels we can
treat in a fully quantum approach.  To circumvent this,
we evaluated $M_{2D}(H)$ at fixed $\Delta_0$ so as to access a large range
of $n_{\mu}$ with the same input order parameter;
this could be physically sensible if the superconducting order were
determined primarily by pieces of Fermi surface much larger than the
one being studied, which is the case for experiments done to date.
These calculations were performed using the
$\lambda = 1 $ model so that the average off-diagonal self-energy
in $\hbar \omega_c$ units is $ F = 0.436 \Delta_0 $ as discussed
previously.  Using the BCS ratio of $F$ to $k_B T_{c2}$ of 1.76,
this would mean that $\Delta_0 \sim 4 $ should give an amplitude
suppression similar to that in the normal state at
$ k_B T = \hbar \omega_c$.   We find that as
$\Delta_0$ is increased from zero to four, peaks in the Fourier
transform of $M(H)$ at all dHvA harmonics are
suppressed, and for values larger than four,
no detectable peaks occur.  (With our convention the fundamental
harmonic of the normal state oscillations occurs in the sine transform;
in our calculations, no signficant effects occur in the cosine transform except
for the constant diamagnetic response).
In Fig. 14, we plot the peak height of the fundamental harmonic
versus $\Delta_0$ and compare it with the result obtained neglecting
Landau level mixing and with the normal state oscillations at the
estimated $T_{c2} \sim 0.248 \hbar \omega_c \Delta_0$.
We see that the suppression of the amplitude of the
oscillations is underestimated by the diagonal approximation at
all values of $\Delta_0$.  We also find that Eq.[\ref{eq:width0}] provides
a perfect fit to the diagonal approximation results.  Comparing the initial
slopes of the depression of the full non-diagonal harmonic to that of the
diagonal one, we infer that the full calculation
has an effective scattering rate about 3.5 times larger than the diagonal one.
Given this, we can infer a scattering rate formula analogous to that of
the diagonal case given in Eq.[\ref{eq:width1}]
\begin{equation}
\hbar \tau_S^{-1}
\sim 2 n_{\mu}^{-1/4} k_B T_{c2}
\label{eq:width3}
\end{equation}
The similarity of normal state oscillations at $T_{c2}$ and
the non-diagonal results also suggests that experiments might be
informatively analyzed by introducing a field-dependent Dingle
temperature due to the superconducting order
which could be compared with $T_{c2}(H)$.  Below, we will compare both of
these approaches to experimental data.  Finally, we caution that the
harmonic structure can be qualitatively altered by the superconducting
order.  This can be seen by the fact that in the non-diagonal calculation,
the harmonic actually changes sign and peaks at $\Delta_0 \sim 2$.  This
behavior is not a phase shift effect, rather it is due to the magnetic
oscillations of the order parameter, which
contribute with opposite sign to the kinetic part as in Fig. 11 (note, it
is the output value of $\Delta_0$ from Eq.[\ref{eq:12}] which enters into the
pairing energy of Eq.[\ref{eq:16}], which
is why these gap oscillations contribute even at fixed input $\Delta_0$).
This contribution is much weaker in the diagonal case, so no sign change
occurs in that case.  It is expected that self-consistency effects will wash
this effect out, so it is doubtful whether such a sign change would be seen
experimentally.  We have found, then, that the fundamental harmonic of the
non-diagonal case can be fit by a term like Eq.[\ref{eq:width0}] with the same
enhancement factor of 3.5 mentioned above (thus justifying
Eq.[\ref{eq:width3}]) plus a `pairing' term which is proportional to
$\Delta_0^2$ times a damping factor (an analogous gap oscillation term
occurs in the work of Maniv {\em et al}\cite{maniv} but with no damping).
Similar effects happen for the higher
harmonics, except that as they are suppressed more rapidly, the sign change
and peak occur at smaller values of $\Delta_0$.  This can lead to an unusual
situation where the second harmonic can dominate the magnetization, as can
be seen in Fig. 15, where we show the $M_{2D}(H)$ data from which these
Fourier transforms were evaluated for the non-diagonal case with $\Delta_0=1$.
This highly unusual behavior might be detectable experimentally,
although we caution that the strong gap oscillation contribution in the
non-diagonal case will be damped additionally by self-consistent effects,
so that
this effect might disappear.  To study these self-consistency effects in more
detail will require going away from a
fully quantum approach like ours to a semiclassical approximation where very
large numbers of Landau levels can be treated.

With the extensive discussion above, we are now able to take a closer look at
the experimental data.  The most complete work to date is that of Harrison
{\em et al} on $\rm{Nb}_3\rm{Sn}$.\cite{spring2}
In this work, they plot the scattering
rate defined in Eq.[\ref{eq:rs}] versus field for three orbits on the Fermi
surface.  As in earlier work, these orbits are fairly small compared to the
Brillouin zone dimensions.  In Fig. 16, we compare the experimental scattering
rate for the one orbit with the most data to three theoretical predictions
(results for the other two orbits are similar).  The first is the diagonal
approximation of Eq.[\ref{eq:width1}], the second the non-diagonal
approximation of Eq.[\ref{eq:width3}], and the third is extracted by equating
a thermal suppression factor with a temperature equal to $T_{c2}$ to
Eq.[\ref{eq:rs}].  $T_{c2}$ was determined using the standard method of
Werthamer {\em et al}.\cite{whh}  As can be seen, Eq.[\ref{eq:width3}] gives
a very good representation of the data, with the diagonal approximation
significantly underestimating the damping and a thermal suppression factor of
$T_{c2}$ strongly overestimating the damping.

\section{Variational Approach}
\label{sec:variational}

In this section we discuss superconductivity in the mixed
state from a variational point of view.  In our formalism the
off-diagonal self-energy (Eq.[\ref{eq:6}]) in the BdG equations
is specified by a discrete set
of parameters, $ \Delta_j$.  As we have mentioned previously,
close to $H_{c2}$ only $\Delta_0$ is
significantly different from zero and the off-diagonal self-energy
can then be characterized by a single parameter.  Given the
off-diagonal self-energy, a BCS-type
variational state can be constructed from the eigenvectors
of the BdG equations.  We can therefore think of the $\Delta_j$
as a set of parameters which specify a trial wavefunction.
The order parameter of this trial wavefunction is
specified by a (in general) different set of parameters which are
evaluated from the solution of the BdG equations using
Eq.[\ref{eq:12}]; in the following we refer to the two sets
of parameters as $\Delta_j^{in}$ and $\Delta_j^{out}$.
The expectation value of the Hamiltonian in this wavefunction
is given by Eq.[\ref{eq:15}] with the pairing energy
evaluated from Eq.[\ref{eq:16}] using $\Delta_j^{out}$.
It is clear from the variational derivation of the
BdG equations that the free energy of the model
has extrema at the $\Delta_j$ which are self-consistent solutions
of the BdG equations.

We will discuss in detail the approximation where we consistently
assume that only $\Delta_0 \ne 0$ so that the superconducting state
is characterized by a single parameter as in the zero-field
situation.  We note that
the off-diagonal self-energy, and therefore the variational state,
depend only on the combination $\lambda \Delta_0^{in} \equiv
\tilde \Delta_0^{in}$.  In Fig. 17 we plot the grand potential
at $k_BT/\mu=10^{-4}$ ($\Omega = E - \mu N$) as a function of the
variational parameter
$ \tilde \Delta_0^{in} $ at
$n_{\mu} =9.5$ and $n_{\mu}=10$ for $\lambda =0.75$, $\lambda = 1.00$,
and $\lambda = 1.25$.  In the normal state, the grand potential reaches
its minimum when the chemical potential is at the mid-point
between Landau levels, {\em i.e.,} when $n_{\mu}$ is half an odd
integer.  At these chemical potential values the normal state
grand potential equals its zero-field value $\Omega_0$.  The
grand potential has a local maximum when $n_{\mu}$ is an integer; at integral
values of $n_{\mu}$ the relative increase in the grand potential
compared to its zero-field value is $1/(2n_{\mu}+1)^2$.  In
Fig. 17 we see that for $\lambda = 0.75$ the free energy minimum
occurs at $\tilde \Delta_0^{in} =0$ for $n_{\mu} =9.5$ and at
$\tilde \Delta_0^{in} \sim 1.8$ for $n_{\mu} = 10$.  Integral
values of $n_{\mu}$ are more favorable for superconductivity
because of the high density of states at the Fermi level.  The
small condensation energy at $n_{\mu} = 10.0 $ reduces the
grand potential difference between the two fields and therefore reduces
the amplitude of the dHvA oscillations.  This is
an alternate point of view on the damping of dHvA
oscillations in the quantum regime.  For the larger values of
$\lambda$ the superconducting condensation energy is either comparable
to or larger than the amplitude of the magnetic oscillations
in the normal state grand potential.  For $\lambda = 1.25$ the
minimum grand potential occurs in the vortex core bound state
regime where magnetic oscillations are essentially absent.
In this regime the grand potential is lower at larger
$n_{\mu}$ (weaker field) as expected because of the diamagnetism
of the mixed state, but the magnetic field dependence is {\it not}
periodic in $n_{\mu}$.

Using this approach we can study the dependence of the equilibrium
order parameter on the model parameter $\lambda$.   From the
self-consistency condition we see that the value of $\lambda$ at
which the self-consistent order parameter value is $\Delta^{out}$ is
\begin{equation}
\lambda = \frac{\tilde \Delta_0^{in}}{\Delta_0^{out} [\tilde
\Delta_0^{in}]}.
\label{eq:lambda}
\end{equation}
In Fig. 18 we plot the order parameter as a function of $\lambda$ for
our model at $n_{\mu} = 40$ and $n_{\mu} = 30$ at $k_B T / \mu =
10^{-2}$.  For $n_{\mu} = 40$ results are also shown for $k_B T / \mu =
10^{-4}$.  We note that the order parameter grows like $ (\lambda -
\lambda_c)^{1/2} $ for $\lambda > \lambda_c$ where $\lambda_c$ is the
coupling strength required for superconductivity to occur.  This
is the expected result within mean-field theory.  $\lambda_c$
decreases with decreasing magnetic field and is expected to
approach zero at zero field since in that limit superconductivity
occurs in this model at arbitrarily small $\lambda$.
For $k_B T / \mu = 10^{-4}$ the inequality $k_B T << \hbar \omega_c$
is satisfied and the quantum regime of strong magnetic field
superconductivity is seen in the persistence of superconducting
order to very small $\lambda $ values.  At half odd integral
values of $n_{\mu}$ the order parameter curve would be
shifted toward larger values of $\lambda$ rather than
smaller values of $\lambda$ in the quantum regime.

In Fig. 19 we compare the dependence of the superconducting
condensation energy $F_{c}$ on the order parameter
with expectations based on Ginzburg-Landau
theory by plotting $F_c/(\Delta_0^{out})^2$ against $(\Delta_0^{out})^2$.
(We define $F_c$ in the grand canonical ensemble as the
difference between the normal state
grand potential and the superconducting state grand potential.)
These results are for $\lambda = 1$; with our simple model
results at other values
of $\lambda$ differ only by a constant vertical shift of the curves.
Plots are shown for $k_B T / \mu = 10^{-2}$ at $n_{\mu} = 40$ and $n_{\mu} =
30$
and for  $k_B T / \mu = 10^{-4}$ at $n_{\mu}=40$ as in Fig. 18.
In Ginzburg-Landau theory\cite{tinkham} these plots should give straight lines
whose $y$ intercepts are proportional to the coefficient of the
quadratic term in the Ginzburg-Landau energy functional for
the lowest Landau level of the Cooper pairs and whose slopes are
proportional to the coefficient of the quartic term in
the Ginzburg-Landau energy functional.  The intercept is expected to
increase with increasing magnetic field and with increasing
temperature in agreement with our calculations.
We see evidence in Fig. 19 for some temperature
and magnetic field dependence in the coefficient of the quartic
term which is expected far away from the zero-field critical
temperature but is normally neglected in Ginzburg-Landau theory.
A large departure from Ginzburg-Landau theory is seen at
small order parameters for the low temperature case.  Again this
deviation appears only in the quantum regime where pairing is
dominantly within an individual Landau level.  At half odd integral
values of $n_{\mu}$ the magnitude of the condensation energy
would be decreased rather than increased upon entering the quantum
regime.

\section{Summary}
\label{sec:conclusion}

In this paper, we have presented an exact mean field treatment of the flux
lattice state and discussed in detail the resulting quasiparticle electronic
structure and magnetization in the 2D limit.  We have found that as the
field is lowered, the Landau levels nearest the chemical potential cross over
to vortex core bound states.  Moreover, the field dependence of the
quasiparticle electronic structure shows rich behavior, due to the strong
field dependence of the Landau levels, order parameter, and vortex-vortex
interactions, and their interplay with one another.
We also find that as the field is lowered below $H_{c2}$,
the single period magnetic oscillations become rapidly damped due to
broadening of the Landau levels, and essentially disappear when
the average superconducting gap becomes of order the cyclotron energy.
This is consistent
with observations of the dHvA effect in the mixed state.  For lower fields,
gaplessness can still occur, but the uniform periodicity of the gaplessness
(as occurs near $H_{c2}$) is destroyed by the various competing
effects mentioned above.  For even lower fields, the vortex core regime is
reached and gaplessness no longer occurs, but there is still structure
in the magnetization driven by the spatial modulation of the order parameter.
We would like to conclude by saying that the rich behavior we predict for
the field dependent quasiparticle electronic structure has important
implications not only for the dHvA effect, but for all experiments which
measure low energy excitations in the vortex lattice state.

\acknowledgments

This work was supported in part by the U.S. Dept. of Energy,
Basic Energy Sciences, under Contract No. W-31-109-ENG-38,
and in part by the National Science Foundation under grant
DMR-9416906.  The authors would like to thank
Prof. Mike Springford for sending us a preprint of the work of his group on
$\rm{Nb}_3\rm{Sn}$ and Profs. Steven Hayden and Michael Naughton for
discussions
concerning experimental data.  M.R.N. would like to acknowledge the hospitality
of the Cavendish Laboratory, Cambridge University, where some of this work
was begun.

\vfill\eject

\begin{figure}
\caption{$T_{c2}$ and the spatial average of the off-diagonal
self-energy $F$ at $k_BT/\mu=10^{-3}$ in units of the chemical potential,
$\mu$,
versus $n_{\mu}$.  These results are for the $\lambda = 0.75 $
model with a smooth cut-off as discussed in the text.  All results
are for this model unless otherwise noted.}
\label{fig1}
\end{figure}

\begin{figure}
\caption{Magnitude of the order parameter ($\Delta$)
at $k_BT/\mu=10^{-3}$ along a line between
neighboring vortices on the triangular vortex lattice.  $r$ is the
distance from one vortex in units of the near-neighbor distance.
The ten curves are for $n_{\mu} =12, 15, \cdots, 39$.
The order parameter is calculated in units of $(4 \pi \ell^2)^{-1}$.
In these units the off-diagonal self-energy is obtained by multiplying
by $\frac{\lambda}{2}\hbar \omega_c$.  As the field is lowered
( $n_{\mu}$ is increased ) the order parameter becomes larger and the
ratio of the vortex core size to the period of the vortex lattice
becomes smaller.}
\label{fig2}
\end{figure}

\begin{figure}
\caption{Lowest ten quasiparticle energies along the perimeter
of the irreducible triangle in $\vec{k}$-space for $n_{\mu}=15$.  The
energies are expressed in units of $\hbar \omega_c$.}
\label{fig3}
\end{figure}

\begin{figure}
\caption{As in Fig.3 but for $n_{\mu}=40$.}
\label{fig4}
\end{figure}

\begin{figure}
\caption{Magnetic field dependence of the
average quasiparticle energies for the lowest five pairs of bands.  The
field is parameterized by $n_{\mu}$ and the quasiparticle energies
are expressed in terms of $\mu$.}
\label{fig5}
\end{figure}

\begin{figure}
\caption{Contributions to the order parameter from the four lowest
energy quasiparticle bands.  The modulus of the order parameter is plotted
along the same line connecting neighboring vortices as in Fig. 2.
These results are for $n_{\mu}=40$.  To aid clarity
we plot the modulus of the contributions
from the first two bands and the negative of the modulus of the
contributions from the second two bands.  We see that the contributions
from each member of a pair are essentially identical.  The contribution
of the first pair of quasiparticle
bands corresponds to what is expected for the lowest
energy bound state in an isolated vortex core while the contribution from the
second pair of bands corresponds to what is expected from the
next bound state in an isolated vortex core.
}
\label{fig6}
\end{figure}

\begin{figure}
\caption{Magnetic field dependence of quasiparticle energies at
21 inequivalent $\vec{k}$ points in the irreducible triangle for the four
lowest bands.  The quasiparticle energies are in units of $\hbar
\omega_c$ and the field is parameterized by $n_{\mu}$.
In these field-dependent units the energy of field-independent eigenvalues is
proportional to $n_{\mu}$.  We see the emergence of weakly field
dependent quasi-bound states at low fields.}
\label{fig7}
\end{figure}

\begin{figure}
\caption{As in Fig.7 but for the model with $\lambda$=0.55.}
\label{fig8}
\end{figure}

\begin{figure}
\caption{Landau level occupations numbers in the mixed state at
$k_BT/\mu=10^{-3}$ for $n_{\mu}=40$.  The solid curve is the Fermi distribution
function evaluated at $T=T_{c2}$ at the same magnetic field.}
\label{fig9}
\end{figure}

\begin{figure}
\caption{$M_{2D}(H)$ for the $\lambda = 0.75$ model
versus $n_{\mu}$.  The magnetization per particle is
in units of $\frac{e\hbar\mu}{m^*c}$.}
\label{fig10}
\end{figure}

\begin{figure}
\caption{Decomposition of $M_{2D}(H)$ into kinetic and pairing energy
contributions for the $\lambda = 0.75$ model.}
\label{fig11}
\end{figure}

\begin{figure}
\caption{$M_{2D}(H)$ versus $n_{\mu}$ for the $\lambda=0.55$ model.}
\label{fig12}
\end{figure}

\begin{figure}
\caption{Quasiparticle Landau level width (standard deviation for 66 $\vec{k}$
points)
at $\xi_{n_{\mu}} = 0 $  ($\hbar \tau_S^{-1}$)
at integral $n_{\mu}$ neglecting inter-Landau level pairing for an input
$\Delta_0=1$ (results including
inter-Landau level terms are very similar).  In this
approximation, the band width is proportional to $\Delta_0$.  The curve
is a fit to a power law of $n_{\mu}^{-1/4}$.}
\label{fig13}
\end{figure}

\begin{figure}
\caption{Plot of the fundamental harmonic of the
sine Fourier transform of $M_{2D}(H)$ over the interval from $n_{\mu}=20$
to $n_{\mu}=40$ as a function of input $\Delta_0$ (solid circles).  The solid
diamonds show
results obtained neglecting Landau level mixing and
the open circles are for the normal state with a temperature equal to
$0.248 \hbar \omega_c \Delta_0$, the estimated $T_{c2}$.  For these
calculations, $\lambda=1$ and $k_BT/\mu=10^{-4}$.}
\label{fig14}
\end{figure}

\begin{figure}
\caption{$M_{2D}(H)$ for an input $\Delta_0=1$ (solid line) in the calculations
of Fig.14.
The dashed lines show results obtained when Landau level mixing is
neglected.}
\label{fig15}
\end{figure}

\begin{figure}
\caption{$\tau_S^{-1}$ in TeraHertz units versus field in Tesla
extracted from the dHvA data
of Ref. 7 for the
581 Tesla orbit of $\rm{Nb}_3\rm{Sn}$ (solid circles)
compared to results of
the diagonal approximation of Eq.[23] (+), including
inter-Landau level effects, Eq.[25] (x), and to a thermal
damping factor equal to $T_{c2}$ (open circles).  Results for other orbits
are similar.  $H_{c2}$ is 19.7 Tesla.}
\label{fig16}
\end{figure}

\begin{figure}
\caption{Dependence of the grand potential on the variational
parameter $\tilde \Delta_0^{in}$ for three different values of
$\lambda$ with $k_BT/\mu=10^{-4}$.  The dashed curves are
for $n_{\mu} = 10$ and the solid curves are for $n_{\mu}=9.5$.}
\label{fig17}
\end{figure}

\begin{figure}
\caption{Equilibrium order parameter as a function of $\lambda$.
The solid line is for $n_{\mu}=30$ and $k_BT/\mu=10^{-2}$; the
dashed line is for $n_{\mu}=40$ and $k_BT/\mu=10^{-2}$; the long-dashed
line is for $n_{\mu}=40$ and $k_BT/\mu=10^{-4}$.}
\label{fig18}
\end{figure}

\begin{figure}
\caption{Superconducting condensation energy as a function of
the order parameter of the variational wavefunction, $\Delta_0^{out}$.
The solid line is for $n_{\mu}=30$ and $k_BT/\mu=10^{-2}$; the
dashed line is for $n_{\mu}=40$ and $k_BT/\mu=10^{-2}$; the long-dashed
line is for $n_{\mu}=40$ and $k_BT/\mu=10^{-4}$.}
\label{fig19}
\end{figure}

\end{document}